\documentclass[transmag]{IEEEtran}
\usepackage{enumerate}
\usepackage{graphicx}
\usepackage{latexsym}
\usepackage{graphicx}
\usepackage{hyperref}
\def\BibTeX{{\rm B\kern-.05em{\sc i\kern-.025em b}\kern-.08em T\kern-.1667em\lower.7ex\hbox{E}\kern-.125emX}}
\usepackage{subfigure}
\usepackage[mathscr]{eucal}
\usepackage{caption}

\usepackage{float}
\usepackage[dvipsnames]{color}
\usepackage{amssymb}
\usepackage[centertags]{amsmath}
\usepackage{newlfont}
\usepackage{epstopdf}
\usepackage{amsthm}
\usepackage{amssymb}
\usepackage{amsmath}
\usepackage{epstopdf}
\usepackage{multirow,url}
\usepackage[sort]{cite}
\usepackage{multicol,lipsum}
\usepackage{mathtools}
\usepackage{cuted}

\usepackage{array}
\usepackage{caption}
\begin{document}
\title{Optical IRS Aided B5G V2V Solution for Road Safety Applications}

\author{\IEEEauthorblockN{ Tathagat Pal, Gurinder Singh, Vivek Ashok Bohara, and Anand Srivastava}\\
Wirocomm Research Group, Department of Electronics \& Communication Engineering\\ Indraprastha Institute of Information Technology, New Delhi, India\\ 
E-mail: \{tathagat19211, gurinders, vivek.b, anand\}@iiitd.ac.in
}
\maketitle

% This research work was supported
% by the Visvesvaraya Ph.D. Scheme by Ministry of Electronics and Information
% Technology (MeitY), Government of India, implemented by Digital India Corporation.

\vspace{-4 cm}

\begin{abstract}
In this work, we showcase the potential benefit of employing  optical intelligent reflecting surfaces (O-IRS) for improving safety message dissemination for vehicular visible light communication (V-VLC) systems particularly at the road intersections. Buildings, roadside structures, signboards, and other impediments commonly hinder line-of-sight (LoS) communication between vehicles at urban crossroads scenarios. We propose using O-IRS at road intersection to improve the communication link's reliability. We compare the performance of proposed scheme with baseline scenarios such as optical relay and non line-of-sight (NLOS) road reflection (NRR) aided vehicle-to-vehicle (V2V) communication. From obtained results, it has been shown that O-IRS offers considerable performance enhancement as compared to the baseline scenarios. In particular, O-IRS can achieve longer communication range as compared to the optical relay aided V-VLC systems while ensuring desired quality-of-service (QoS).
\end{abstract}

\begin{IEEEkeywords}
 Vehicular visible light communication (V-VLC), vehicle-to-vehicle (V2V), intelligent reflective surfaces (IRS), optical relay, NLOS road reflections (NRR).
 \end{IEEEkeywords}
 
\section{Introduction}
Every year, around 1.35 million people die as a result of a traffic accident. As a consequence, individuals and their families, and the country as a whole suffer massive financial losses, which costs most countries 3\% of their GDP \cite{who}. To address this global problem, researchers are constantly working towards developing more advanced features enabled intelligent transportation systems (ITS).  There are  three major types of communications in a vehicular network namely: infrastructure to vehicle (I2V), vehicle to infrastructure (V2I), and vehicle to vehicle (V2V) communications. Vehicle-to-everything (V2X) communication, a key enabler for ITS, allows opportunistic information exchange among vehicles as well as with road side unit (RSU) which can provide valuable assistance in minimizing the road accidents \cite{liu20206g}. IEEE 802.11p and cellular-V2X (C-V2X) are two RF-based communication technologies that have been designed particularly for vehicular adhoc networks (VANETs). However, the RF technologies discussed above may not always be capable of handling all types of application scenarios that need high throughput and low latency (for instance, vehicular platooning). The task is aggravated by the restricted RF spectrum which might not be quite enough to handle the rising demands for future ITS.

Against the above background, visible light communication (VLC) offers a cost-effective way to establish V2X communications. Vehicular visible light communication (V-VLC) utilizes light
emitting diode (LED) equipped vehicle's headlamp as transmitter and photodetector
(PD) or camera image sensors as the receivers, thereby exhibiting dual
benefit of vehicular light: high data rate communication and illumination simultaneously. The significant attributes of V-VLC such as lower
power consumption, high data rate, reduced transceiver design cost, enhanced security, enhanced link quality and anti-electromagnetic interference makes it a potential candidate for beyond 5G (B5G) V2X networks \cite{liu20206g}. The performance of V-VLC is substantially confined by its direct line-of-sight (LOS) necessity in realistic situations or instances wherein LOS path may not be present.

Recently, intelligent reflecting surfaces (IRS) have emerged as a possible alternative for green and cost-effective communication technology to improve the signal quality and transmission coverage
via passive reflecting arrays. In \cite{basar2019wireless}, the authors present a comprehensive analysis and historical perspective on current IRS-assisted communication system solutions. In the context of VLC systems, IRSs are expected to play a crucial role in enhancing their performance, given that the majority of VLC systems rely on the availability of LOS link. An IRS consists of numerous passive antenna elements having reconfigurable phases. For VLC systems, there are two types of optical-IRS (O-IRS) namely, metasurface reflectors and intelligent mirror arrays. In \cite{9276478}, the authors have proposed the integration of programmable metasurfaces and mirror arrays based IRS with an indoor VLC systems. For an outdoor environment, the integration of IRS in RF based vehicular communication networks have been well studied and illustrated  in \cite{qi2020traffic,chen2020resource,makarfi2020reconfigurable}. 

 \begin{figure*}[t]
\centering
  \subfigure[A typical road intersection scenario where vehicles
in blocked line-of-sight (solid red line) communicate via
 O-IRS (solid green line) in V-VLC systems.]{%
  \includegraphics[width=0.47\textwidth,height=5cm]{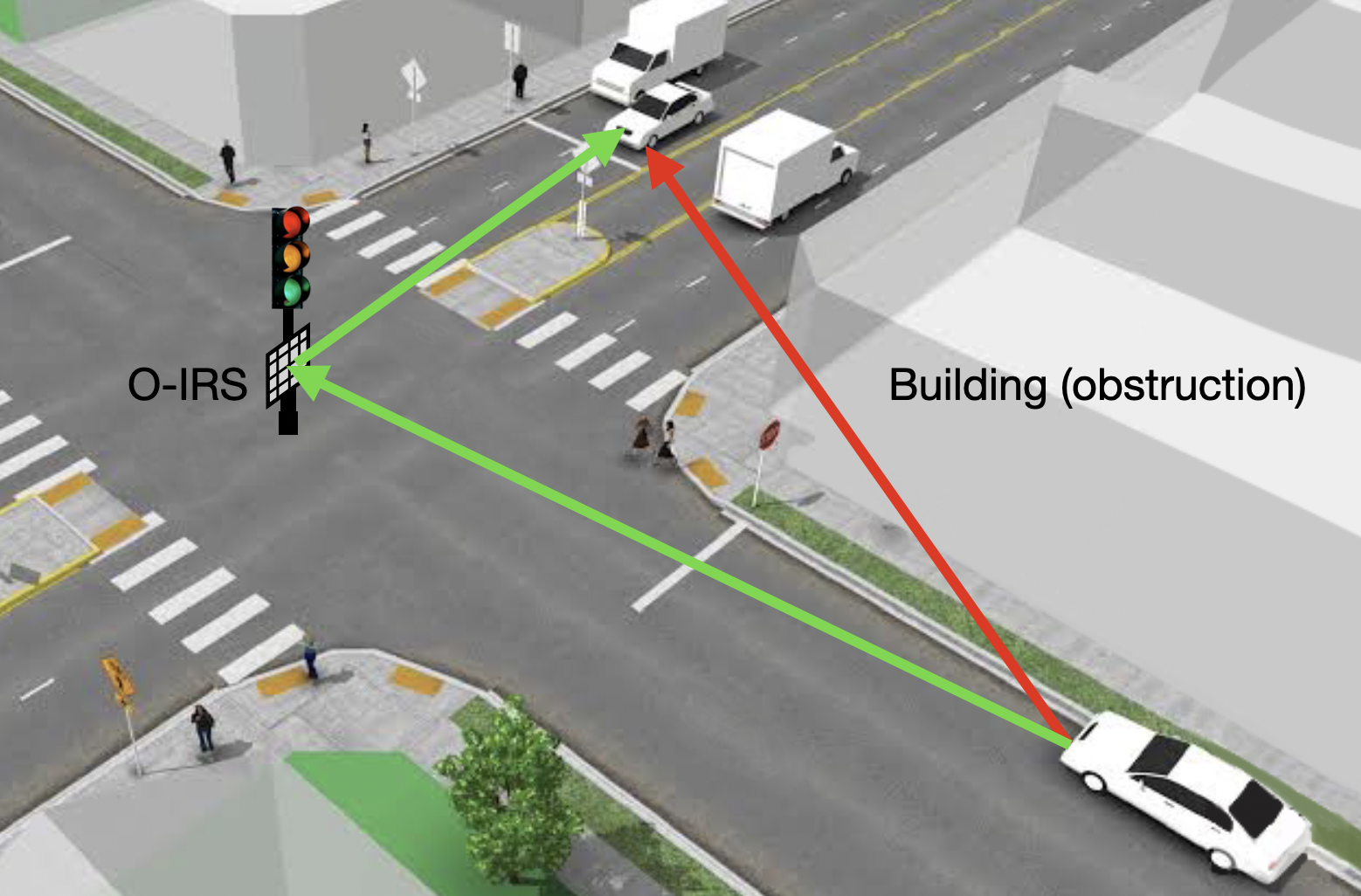}
 \label{fig_rls}}
 \subfigure[O-IRSs can be deployed on road surface to relax the LOS requirement between source and destination vehicles in V-VLC systems.]{%
 \includegraphics[width=0.47\textwidth,height=5cm]{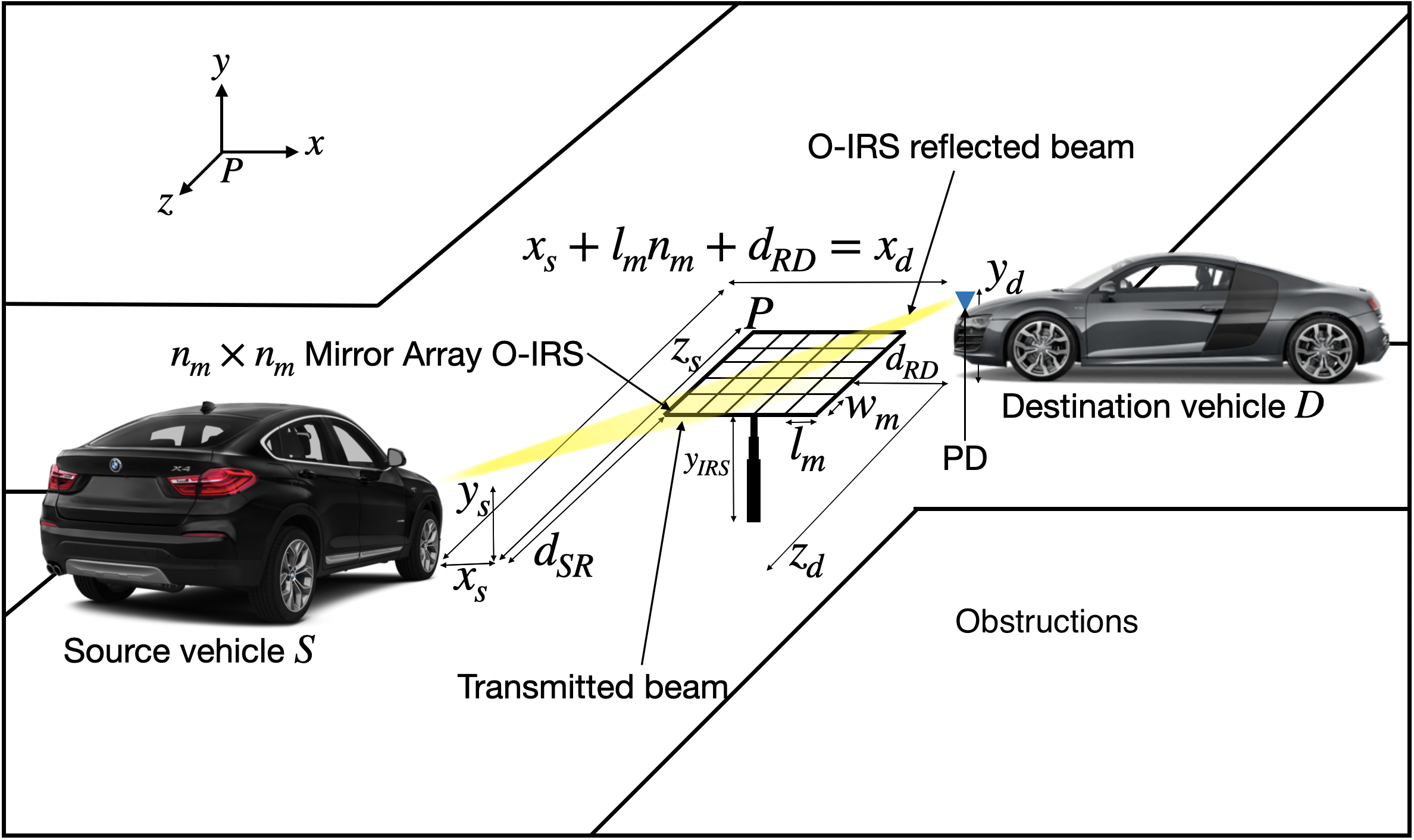}
 \label{fig_IRS}}
 \caption{Proposed O-IRS scheme: (a) Real life scenario and (b) 3D-coordinate system model for O-IRS aided V2V.}
 \label{fig9}
 \end{figure*}

To the best of authors' knowledge, the proposed research study is first of its kind to pioneer the use of optical IRS (O-IRS) with V-VLC for road safety applications in vehicular environments. The proposed scheme could help solve the NLOS transmission problem in a V-VLC systems for road intersections, where significant amount of road crashes usually occur. In urban environments, particularly at road crossings, where the opportunistic transmission of safety messages between vehicles by visible light might be obstructed by vehicles in adjacent lanes, edifices, and more such impediments, O-IRS is an effective use case. Furthermore, vehicles in perpendicular lanes may be unable to communicate effectively, leading to reduction in the performance of urban V2V communication. We assess the efficacy of our suggested model by contrasting it to the standard baseline scenarios where we consider optical relay aided V2V and NLOS road reflections (NRR) aided V2V communications.  

The remainder of this paper is organized as follows: Section II presents detailed system model for V2V communication via an O-IRS, optical relay and NLOS road reflections (NRR. Section III discusses the numerical results. Finally, the concluding remarks have been made in Section IV. 

\section{System Model}
We examine a road intersection case as depicted in
Fig. \ref{fig_rls}. For V-VLC systems,
 the direct LOS link from a
vehicular node to the desired receiver may be blocked by obstructions
such as buildings and road side installations. In such case, the communication between vehicles can be facilitated by an O-IRS. In the following, we discuss the performance of O-IRS aided V2V and other baseline scenarios such optical relay and NRR aided V2V communication for a road intersection scenario. For ease of analysis we assume that the vehicle tends to drive slowly at the road intersection which resembles a quasi-static vehicular environment \cite{9772417}.

% \begin{figure}
% \centerline{\includegraphics[width=0.5\textwidth,height=6cm]{RLS.png}}
% \caption{Illustration of road intersection scenario where vehicles
% in blocked line-of-sight (solid red line) communicate via
%  O-IRS/PMS (solid green line) in V-VLC systems.}
% \label{fig_rls}
% \end{figure}

% \begin{figure}
% \centerline{\includegraphics[width=0.5\textwidth,height=6cm]{v2v_irs_modell.png}}
% \caption{O-IRSs can be deployed on road surface to relax the LOS requirement between source and destination vehicles in V-VLC systems.}
% \label{fig_rls}
% \end{figure}
 
 \begin{figure*}[t]
\centering
  \subfigure[Optical relay aided V2V communication system.]{%
  \includegraphics[width=0.46\textwidth,height=5cm]{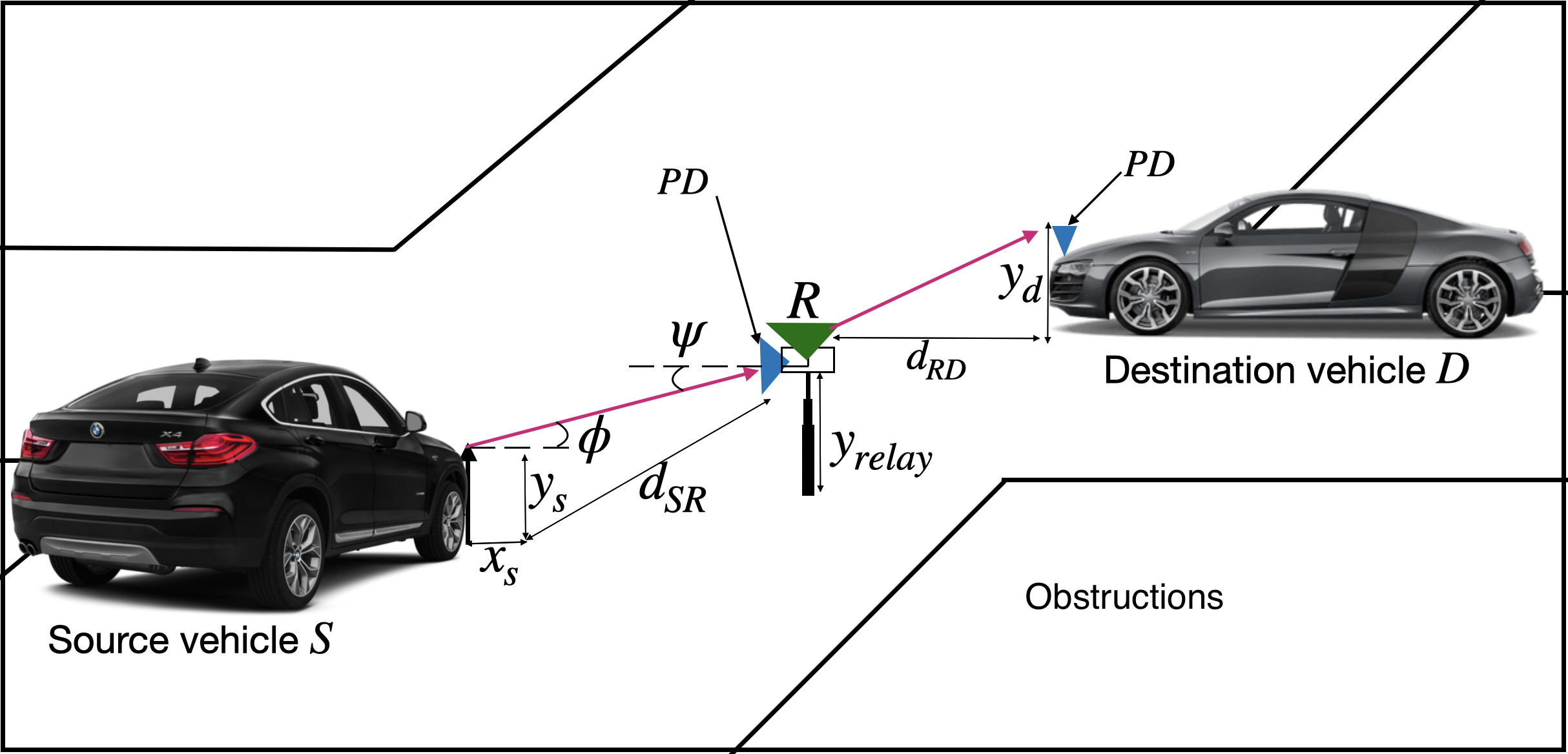}
 \label{relay}}
 \subfigure[Potential of considering NLOS Road reflection (NRR) for vehicular communication. Road surfaces can realize anomalous reflection in a desired vehicle's direction.]{%
 \includegraphics[width=0.45\textwidth,height=5cm]{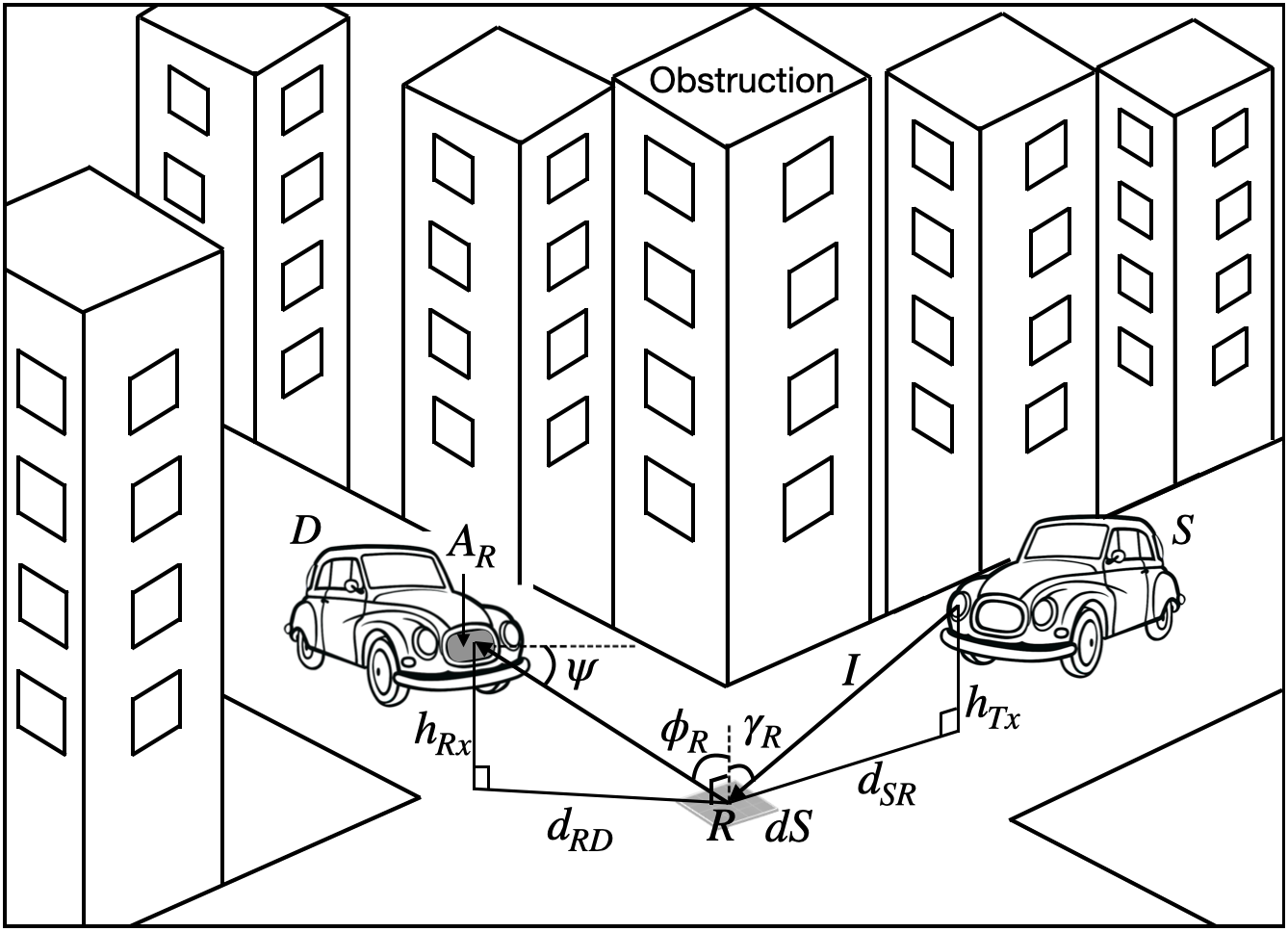}
 \label{fig_PMS}}
 \caption{3D-coordinate system model: (a) Optical relay aided V2V and (b) NRR aided V2V.}
 \label{fig91}
 \end{figure*}

\vspace{-0.238 cm}

\subsection{O-IRS Aided V2V}
 O-IRS can be implemented by using meta surfaces or mirror array which are programmed via an O-IRS controller. Based on the insights given in \cite{9276478}, use of mirror array O-IRS performs more effectively than meta-surface O-IRS for a VLC system. As a result, for this work, we take into account the mirror array O-IRS. Fig. \ref{fig_IRS} illustrates O-IRS aided V2V communication where the mirror array O-IRS consisting of $n_m \times n_m$ identical mirror elements is mounted on a road side unit (RSU). The length and width the mirror element is $l_m$ and $w_m$, respectively. Each mirror element can be rotated around $x$-axis ($\alpha_{i,j}$) and around $z$-axis ($\beta_{i,j}$). The channel gain $G_{IRS}$ of the reflected signal from intelligent mirror array O-IRS is given as follows \cite{9276478}:
\begin{equation}\label{h_irs}
    G_{IRS}(\alpha, \beta)= \sum_{i=1,j=1}^{{all \,mirrors}} \eta A_R T g(\theta_{\textbf{R}_{i,j}}^\textbf{D}) \mathcal{I}_{i,j}(\alpha,\beta)  \mbox{cos}(\theta_{\textbf{R}_{i,j}}^\textbf{D}),
\end{equation}
where $\eta$ is the current-to-light conversion efficiency of the LED, $A_R$ is the area of the receiver, $T$ is the optical filter gain. In \eqref{h_irs}, $g(\theta_{\textbf{R}_{i,j}}^\textbf{D})$ denotes the gain of an optical concentrator and is expressed as: 

\begin{equation}
g(\theta_{\textbf{R}_{i,j}}^\textbf{D}) = \left\{
        \begin{array}{ll}
            \frac{n^2}{\mathrm{sin^2(\Psi_{c})}},  & \quad 0\leq \theta_{\textbf{R}_{i,j}}^\textbf{D}\leq\Psi_{c}, \\
            0 & \quad \theta_{\textbf{R}_{i,j}}^\textbf{D} > \Psi_{c},
        \end{array}
    \right.
\end{equation}
where $n$ is the refractive index of optical concentrator and $\psi_{c}$ is half the receiver's field of view (FOV). 
 
Further, $\mathcal{I}_{i,j}$ represents the irradiance at the receiver contributed by the $(i, \,j)^{th}$ mirror element. $\mathcal{I}_{i,j}$ is given as \cite{9276478}:
\begin{equation}\label{I_irs}
\begin{aligned}
\mathcal{I}_{i,j}(\alpha,\beta)=&\frac{(m+1)\rho_M}{2\pi}\int_{\frac{-l_m}{2}}^{\frac{l_m}{2}}\int_{\frac{-w_m}{2}}^{\frac{w_m}{2}}cos^{m}(\theta^\textbf{I}_{\textbf{R}_{i,j}}) \\
     & \times \frac{\textbf{e}^T_3(\textbf{D} - \textbf{R}_{i,j}){\hat{\textbf{N}}}^T_{i,j}(\textbf{D}-\textbf{R}_{i,j})}{\|\textbf{D}-\textbf{R}_{i,j}\|^4_2} \\
     & \times\mathbb{I}\Big(\textbf{e}^T_1\textbf{S}-\frac{w_s}{2}\leq \textbf{e}^{T}_1\textbf{I} \leq \textbf{e}^T_1\textbf{S} + \frac{w_s}{2}, \\
     & \textbf{e}^T_2\textbf{S}-\frac{l_s}{2}\leq\textbf{e}^T_2\textbf{I}\leq \textbf{e}^T_2\textbf{S}+\frac{l_s}{2}\Big)dx''dz'',
\end{aligned}
\end{equation}

% \begin{figure}
% \centerline{\includegraphics[width=0.5\textwidth]{v2v_irs.png}}
% \caption{O-IRSs can be deployed on road surface to relax the LOS requirement between source and destination vehicles in V-VLC systems.}
% \label{fig_IRS}
% \end{figure}

% \begin{equation} \label{irradiance}
%     E_{i,j} = n_m^2\frac{\rho(m+1)\mbox{cos}^m(\theta_{\textbf{R}_{C}}^\textbf{S})\mbox{cos}(\theta_{\textbf{R}_{C}}^\textbf{D})}{2\pi(\|\textbf{R}_{C}\textbf{D}\|_2 + \|\textbf{R}_{C}\textbf{S}\|_2)^2}
% \end{equation}
 where, $m$ represents the Lambertian order which is given by
$m = \frac{-\mbox{ln}(2)}{\mbox{ln}(\mbox{cos}(\Phi_{\frac{1}{2}}))}$, $\phi_{\frac{1}{2}}$ is the LED semi-angle, $\rho_M$ is the reflection efficiency of mirror, $\mathbb{I(.)}$ denotes the binary indicator function,  .

The vectors $\textbf{S}$ (the source coordinates of the center of the source) and $\textbf{R}_{i,j}$ (the centroid coordinates of the mirror in $i^{th}$ row and $j^{th}$ column) are given as:
 \begin{equation} \label{eq7}
    \textbf{S} = \begin{bmatrix}
          -\Big(x_s + \frac{w_m}{2}+ (j-1)w_m\Big)\\
          -(y_{IRS}-y_s) \\
          -\Big(z_s + \frac{l_m}{2}+ (i-1)l_m\Big)
         \end{bmatrix},
\end{equation}

 \begin{equation} \label{eq8}
    \textbf{R}_{i,j} = \begin{bmatrix}
           \Big(x_s + \frac{w_m}{2}+ (j-1)w_m\Big) \\
           0 \\
           \Big(z_s + \frac{l_m}{2}+ (i-1)l_m\Big)
         \end{bmatrix},
\end{equation}

where $x_s$ and $z_s$ denotes the distances between $S$ and $P$ along the direction of $x$ and $z$-axis respectively, $y_{IRS}$ is the height of the RSU as depicted in Fig. \ref{fig_IRS}. Observe that the distance of the source vehicle, $d_{SR} = z_s - n_mw_m$, and the height of the transmitter (headlights), $h_{Tx} = y_s$. 

Similarly, $\textbf{D}$ (the destination coordinates) can be obtained as:

\begin{equation} \label{eq9}
    \textbf{D} = \begin{bmatrix}
           x_d-\Big(x_s + \frac{w_m}{2}+ (j-1)w_m\Big) \\
           -(y_d-y_{IRS}) \\
           z_d-\Big(z_s + \frac{l_m}{2}+ (i-1)l_m\Big)
         \end{bmatrix},
\end{equation}

$x_d$ and $z_d$ are the distances between \textbf{$\textbf{D}$} and \textbf{$\textbf{S}$} along the direction of $x$ and $z$-axis respectively, as depicted in Fig. \ref{fig_IRS}. $y_d$ is the height of the receiver ($h_{Rx}$). The distance of the destination vehicle, $d_{RD} = x_d - (x_s + n_ml_m)$.

$\textbf{I}$ is the pre-reflection image of $\textbf{D}$ in the source plane (X-Y plane), and is given as:
 \begin{equation} \label{eq81}
    \textbf{I} = \begin{bmatrix}
           \textbf{e}^T_1 \Big(\textbf{R}_{i,j} + \frac{\textbf{e}^T_3(\textbf{S}-\textbf{R}_{i,j})}{\textbf{e}^T_3\widehat{\textbf{R}_{i,j}\textbf{I}}} \widehat{\textbf{R}_{i,j}\textbf{I}} \Big) \\
           \textbf{e}^T_2 \Big(\textbf{R}_{i,j} + \frac{\textbf{e}^T_3(\textbf{S}-\textbf{R}_{i,j})}{\textbf{e}^T_3\widehat{\textbf{R}_{i,j}\textbf{I}}} \widehat{\textbf{R}_{i,j}\textbf{I}} \Big) \\
           \textbf{e}^T_3\textbf{S}
         \end{bmatrix},
\end{equation}

where $\textbf{e}_c$ represents the $c^{th}$ column of an identity matrix of size $3 \times 3$, and $(.)^T$ denotes the transpose function. Furthermore, using $\theta'$ which is the angle between $\textbf{N}_{i,j}$ and reflected light from $\textbf{R}_{i,j}$ to $\textbf{D}$ (cos$^{-1}\big(\widehat{\textbf{N}^T}_{i,j}(\textbf{D}-\textbf{R}_{i,j})/\|(\textbf{D}-\textbf{R}_{i,j})\|_2\big)$),   $\widehat{\textbf{R}_{i,j}}$ can be calculated as $2 \mbox{cos}(\theta')\widehat{\textbf{N}}_{i,j}-(\textbf{D}-\textbf{R}_{i,j})/\|(\textbf{D}-\textbf{R}_{i,j})\|_2$. Here, $\|.\|_2$ denotes the $l_2$ norm of a vector.

The mirror orientation is configured such that $\widehat{\textbf{SR}}_{i,j}$ typifies the corresponding incidence direction for the reflection direction $\widehat{\textbf{R}_{i,j}\textbf{D}}$. Finally, we find a normal vector to the mirror's surface, $\widehat{\textbf{N}}_{i,j}$, to determine its orientation. The vector $\widehat{\textbf{N}}_{i,j}$ can be expressed as \cite{9276478}: 
\begin{equation}\label{eq10}
    \widehat{\textbf{N}}_{i,j} = \frac{\widehat{\textbf{R}_{i,j}\textbf{S}} + \widehat{\textbf{R}_{i,j}\textbf{D}}}{\sqrt{2 + 2 \widehat{\textbf{R}_{i,j}\textbf{S}}^T \textbf{R}_{i,j}\textbf{D}}}.
\end{equation}

The angles defining the mirror orientation ($\beta_{i,j}$ and $\alpha_{i,j}$) can be written using $\widehat{\textbf{N}}_{i,j}$ as follows:
\begin{equation}\label{eq11}
    \beta_{i,j} = \mbox{sin}^{-1}(\widehat{\textbf{N}}^T_{i,j} \textbf{e}_3),\,\mbox{and}\, \alpha_{i,j} = \frac{\mbox{sin}^{-1}(\widehat{\textbf{N}}^T_{i,j} \textbf{e}_1)}{\mbox{cos}(\beta_{i,j})}.
\end{equation}

% and, 
% \begin{equation}\label{eq12}
%     \alpha_{i,j} = \frac{\mbox{sin}^{-1}(\widehat{\textbf{N}}_{i,j} \textbf{e}_1)}{\mbox{cos}(\beta_{i,j})}.
% \end{equation}

% For V2V communications, the optical source (car headlamp) dimension is negligible as compared to the distance between the O-IRS and the source vehicle, ($d_{SR}$). As a consequence, any point on the O-IRS observes all source points experiencing the same location, which allows source to be considered as a point source. 

Finally, $\theta_{\textbf{R}_{i,j}}^\textbf{D}$ denotes the angle between ${\textbf{R}_{i,j}}\textbf{D}$ and the positive $z$-axis and can be obtained as $\textbf{e}^T_3(\textbf{R}_{i,j}-\textbf{I})/\|\textbf{R}_{i,j}-\textbf{D}\|_2$ \cite{9276478}. 

Given the transmitted signal $x$, the received signal $y$ is given as:
\begin{equation}\label{p_irs}
    y=\mathcal{R}G_{IRS}x+w,
\end{equation}
where $\mathcal{R}$ is the responsivity of the PD, $w$ is additive white Gaussian noise (AWGN) with mean zero and variance, $\sigma^2$. With on-off keying (OOK) as modulation scheme\footnote{OOK modulation is one of the standard modulation schemes defined in the VLC standard (IEEE
802.15.7) \cite{rajagopal2012ieee}.}, the bit error rate (BER) can be expressed as:
\begin{equation}\label{ber}
    \mbox{BER} =  \mathcal{Q}\Big(\sqrt{\frac{(\mathcal{R}G_{IRS})^2P_{IRS}}{\sigma^2}}\Big),
\end{equation}
where  $P_{IRS}$ is the transmit power and $\mathcal{Q}(.)$ is the Q-function. 

\subsection{Optical Relay Aided V2V}
The use of optical relays in vehicular communication can effectively
turn a single NLOS link into multiple LOS links. Fig. \ref{relay} portrays optical relay assisted V2V communication. We employ amplify and forward (AF) relay scheme between the source vehicle, $S$ and the destination vehicle, $D$. For a V-VLC system, the DC channel gain $G_{SR}$ can be calculated using \cite{vlc}:

\begin{equation}\label{eq_hsr}
        G_{SR}=
        \left\{ \begin{array}{ll}
            \frac{(m+1)A_R \cos^m(\phi)}{2\pi d_{SR}^2}  T(\psi) g(\psi)\cos(\psi), & 0\leq\psi\leq\psi_{c} \\
            0, &\psi>\psi_c
        \end{array} \right.
\end{equation}

where $m$ represents Lambertian order,
$A_R$ is the area of the photodetector at the receiver, $\psi$ and $\phi$ are the the angle of irradiance and the angle of incidence, respectively. $T(\psi)$ is the non-imaging concentrator. The gain of an optical concentrator, $g(\psi)$ can be expressed as $g(\psi)= \frac{n^2}{\mathrm{sin^2(\psi_{c})}}, \;  0\leq\psi\leq\psi_{c}$, where $\psi_c$ is the field of view (FOV) of the PD, and $d_{SR}$ is the distance between the source vehicle $S$ and the relay $R$. The received power, $P_R$ at the relay can be expressed as:
\begin{equation}
    P_R = (\mathcal{R}G_{SR})^2{P_{SR}},
\end{equation}

where $P_{SR}$ is the transmit power of the source vehicle $S$. Since AF relay scheme is  employed at $R$, $R$ amplifies the signal received from $S$ and forwards it to $D$ \footnote{For sake of fair comparison, we consider that the transmit power at both $S$ and $R$ to be half the transmit power of O-IRS system in order to ensure that the total power budget of both the systems are equal. }. The received signal at $D$ can be expressed as \cite{8659786}:
\begin{equation}
    y_D = G H_{RD}\sqrt{\frac{1}{d_{D}^\mu}}(P_{SR} G_{SR}x+w_1) + w_2,
\end{equation}
where the normalized power $G = \sqrt{\frac{P_{RD}}{P_R + \sigma^2}}$, $\mu$ is the path-loss exponent and $d_D = \sqrt{(y_{relay}-y_D)^2 + d_{RD}^2}$, $w_1$ and $w_2$ are AWGN modelled as $\mathcal{N}$(0, $\sigma^2$). 

The received power at destination vehicle, $D$ can be calculated using \cite{8659786}:
\begin{equation} \label{power_RD}
    P_{D} = P_{RD} (\mathcal{R} H_{RD})^2 P_R.
\end{equation}

The received SNR under AF relay scheme can be approximated as \cite{8659786}:
\begin{equation}\label{snr_af}
    \gamma_{AF} \approx \frac{P_{RD} (\mathcal{R} H_{RD})^2 P_R}{P_{RD} (\mathcal{R} H_{RD})^2 \sigma^2 + d_D^\mu(P_R + \sigma^2)\sigma^2}
\end{equation}

Assuming OOK modulation, \eqref{snr_af} can be used to calculate BER using \eqref{ber}.

\subsection{NLOS Road Reflections (NRR) Aided V2V}
In a V-VLC system, the road surface area
between a transmitter and receiver provides the strongest NLOS components whose strength typically depends on reflection characteristics of road surface. In general, the reflection properties of road surfaces are complex to describe. It can be modeled using luminance coefficients for range of angles, which has been developed for different road surface classifications based on large number of photo metric measurements. For NRR aided V2V, the received power, $dP_{Rx}$ from single reflected path via point $R$ with small area, $dS$ on the road surface at $D$ can be expressed as \cite{c2c}:
\begin{equation} \label{eq_power}
\begin{split}
    dP_{Rx} = \frac{I(\zeta_R,\,\xi_R) \, \mbox{sin}\,\gamma_R \, \rho_R \mbox{cos}\phi_P A_R \, \mbox{cos}\psi\;dS}{\mbox{LER}. \pi .d^2_{\;SR}.d^2_{\;RD}}. 
\end{split}
\end{equation}

where $I(\zeta_R,\,\xi_R)$, LER, $dS$, $A_R$, and $R(\phi_R)$ denote the luminous intensity (in candela) of car headlamp from direction $(\zeta_R,\,\xi_R)$, the luminous efficacy of radiation, a small area on the road surface, the area of receiver, and the reflected radiant intensity respectively. Further, $d_{SR}$, $d_{RD}$, $\gamma_R$, $\phi_R$ and $\psi$ have their usual meaning \cite{c2c} and are also illustrated in Fig. \ref{fig_PMS}.
% where $I$ is the luminous intensity (in candela), $\gamma_R$ is the angle between the normal to the point $R$ on the road surface and the incident signal and $d_{SR}$ is the distance between the light source and $R$, as shown in Fig. \ref{fig_PMS}. LER is the luminous efficacy of radiation, $dS$ is a small area on the road surface. $A_R$ and $d_{RD}$ are the area of receiver and distance between $P$ and $D$, respectively. $R(\phi_R)$ is the reflected radiant intensity \cite{c2c}, $\rho_R$ denotes reflection efficiency, $\phi_R$ denotes the polar angle of scattered light from point $P$ to destination and $\psi$ is angle of incidence of reflected link from view of PD as shown in Fig. \ref{fig_PMS}.

\begin{table}[t]
\centering
\caption{Simulation Parameters}
\label{tab:chn}
\begin{tabular}{|c|c|}
\hline
\textbf{Parameter} & \textbf{Value} \\ \hline
Length of IRS element, $l_m$ & 0.01 m \\  
\hline
Width of IRS element, $w_m$ & 0.01 m \\ 
\hline
Source's $x$-coordinate, $x_s$ & -1.75 \,m \\ 
\hline
Height of Tx vehicle, $h_{Tx}$ or $y_s$  & 0.8 m \cite{c2c} \\
\hline
Luminous Efficacy of Radiation, LER & 150 lm/W \cite{c2c} \\
\hline
Receiver's area, $A_R$ & 1 $\mbox{cm}^2$ \\
\hline
Height of Rx vehicle, $h_{Rx}$ or $y_d$ & 0.8 m \cite{c2c}\\
\hline
Transmission Power for O-IRS System, $P_{IRS}$ & 20 W \cite{9276478} \\
\hline
Transmit Power of $S$ and $R$, $P_{SR} = P_{RD}$ & 10 W \\
\hline
Responsivity of PD, $\mathcal{R}$ & 0.54 A/W \\
\hline
Lambertian order, $m$ & 2 \\
\hline
Optical filter gain, $T$ & 1 \\
\hline
FOV of receiver (FOV) & 60\textdegree \\
\hline
Path loss exponent, $\mu$ & 2 \\
\hline
Noise variance, $\sigma^2$ & $10^{-22}$W \\
\hline
Reflection efficiency of mirror, $\rho_M$ & 0.8  \cite{9276478}\\
\hline
Reflection efficiency of road surface, $\rho_R$ & 0.24 \cite{rho}\\
\hline
Intensity of car headlight, $I(\zeta,\xi)$ & 60000 cd \cite{intensity} \\
\hline
\end{tabular}
\end{table}

\begin{figure*}[t]
\centering
  \subfigure[Received power versus $d_{SR}$ for different values of $n_m$.]{%
  \includegraphics[width=0.46\textwidth]{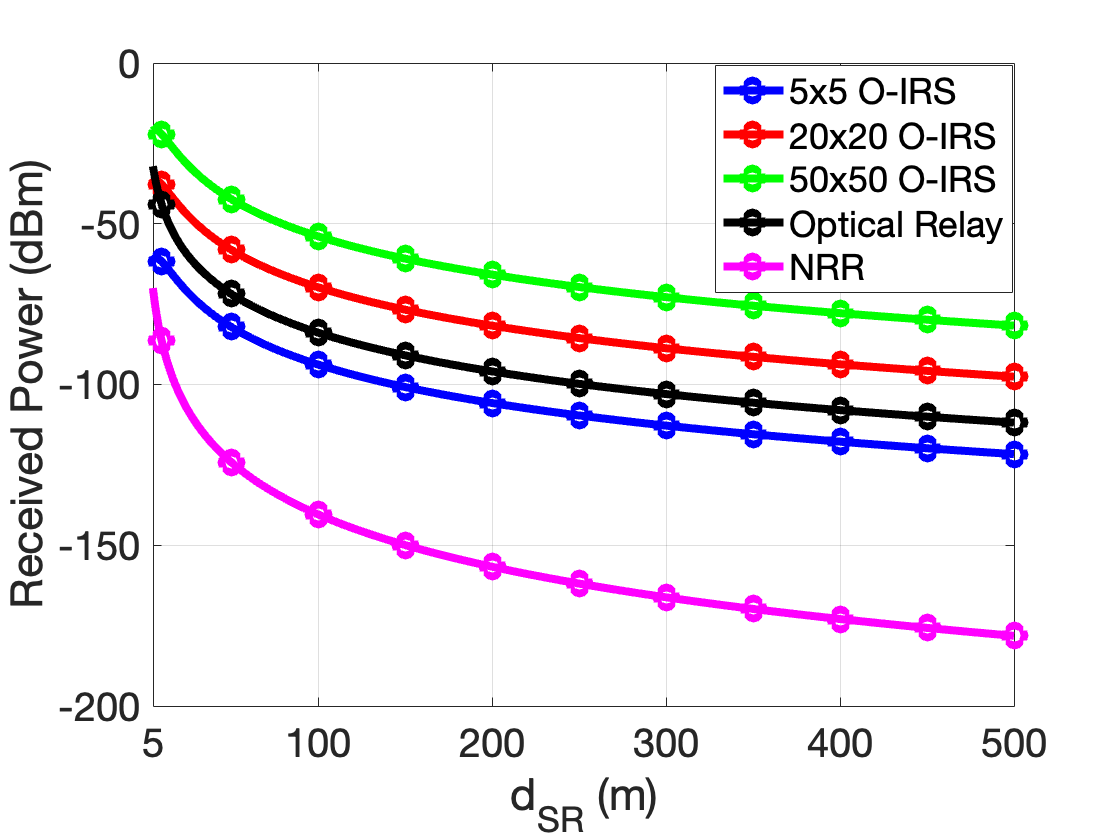}
 \label{fig_power_dist}}
 \subfigure[Impact of $n_m$ on received power, for $d_{SR} = 100 \,\mbox{and} \,200\,\mbox{m}$.]{%
 \includegraphics[width=0.45\textwidth]{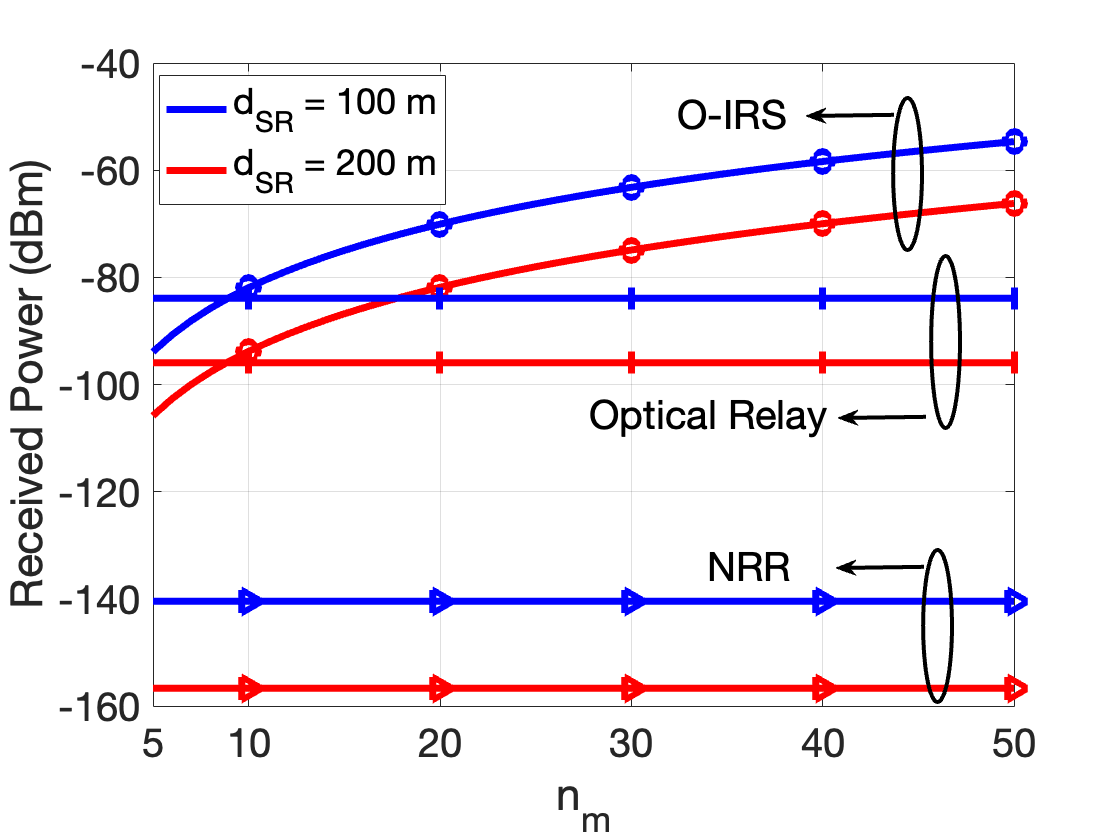}
 \label{fig_power_nm}}
 \caption{Received power of the proposed scheme and the baseline scenarios.}
 \label{fig_power}
 \end{figure*}

\begin{figure*}[t]
\centering
  \subfigure[]{%
  \includegraphics[width=0.46\textwidth]{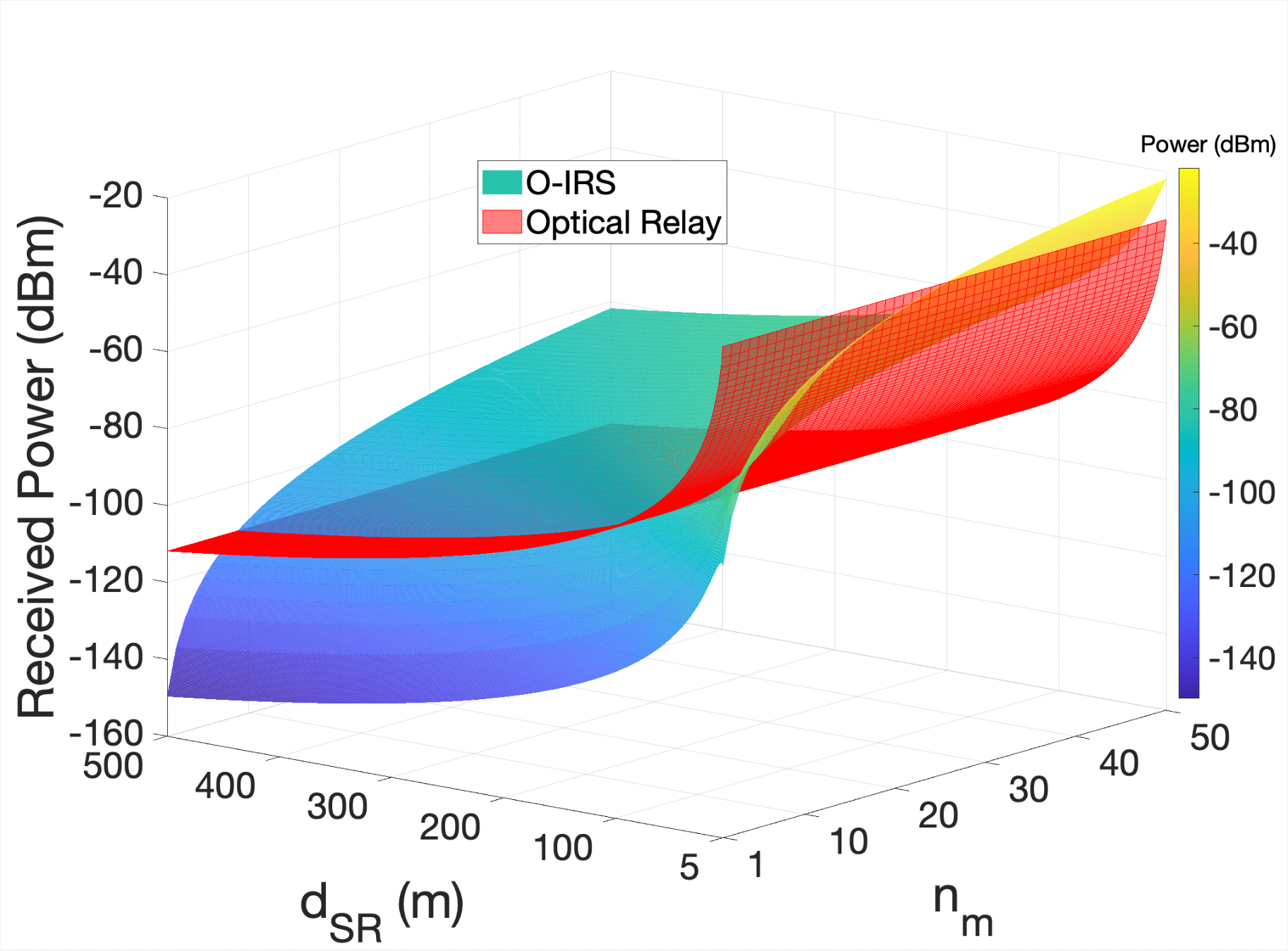}
 \label{fig_3d}}
 \subfigure[]{%
 \includegraphics[width=0.45\textwidth]{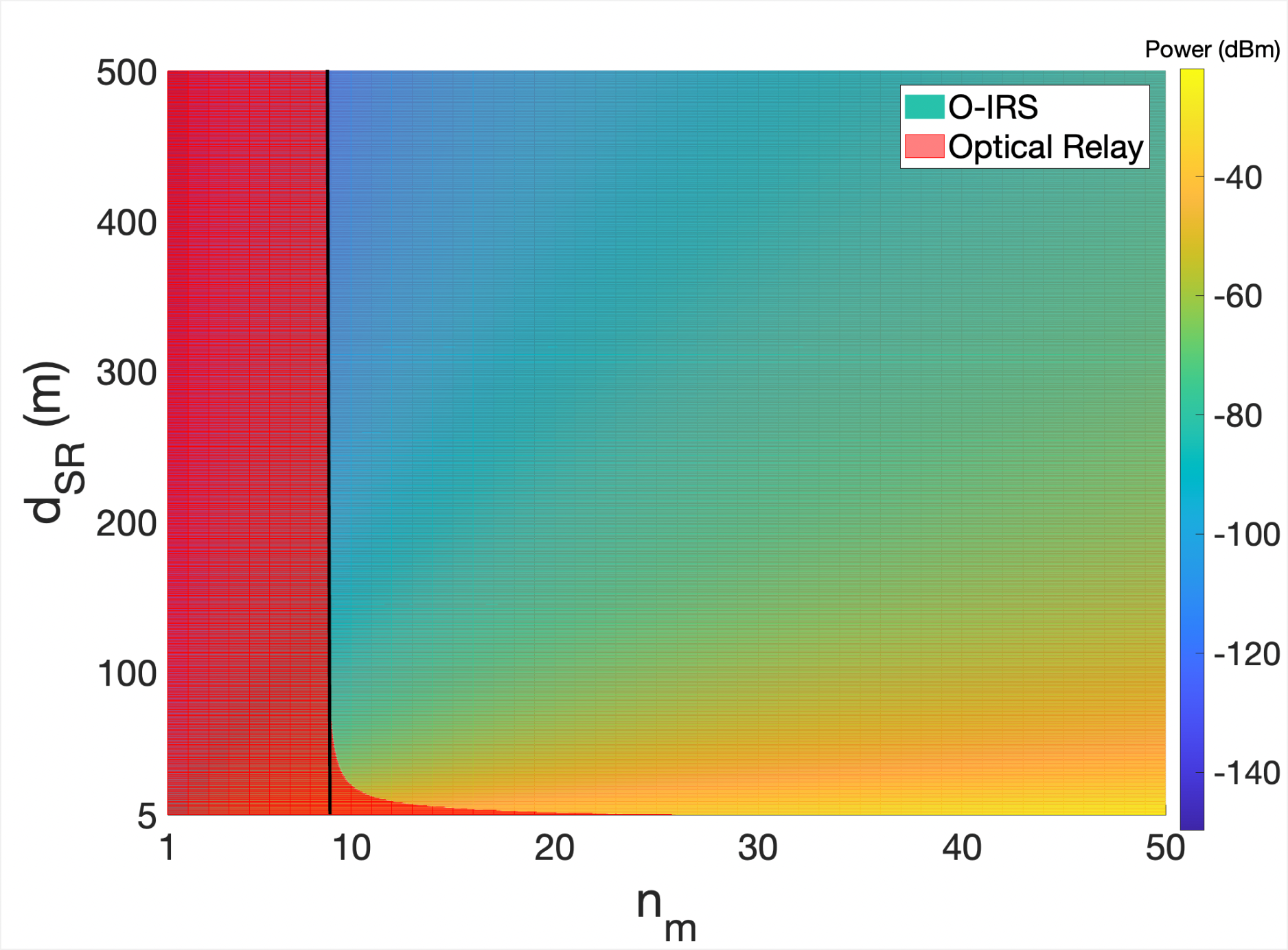}
 \label{fig_topview}}
 \caption{Received power comparison for proposed O-IRS scheme and optical relay aided V2V with varying $d_{SR}$ and $n_m$: (a) illustrates 3D power plots for O-IRS and optical relay aided V2V communication systems, and (b) Top view illustrating 2D $d_{SR}$-$n_m$ curve.}
 \label{fig_3d_ab}
 \end{figure*}

Based on geometrical layout depicted in Fig. \ref{fig_PMS}, it can be readily verified that $\mbox{cos}\psi = \frac{d_{RD}}{\sqrt{(d_{RD}^2+h_{Rx}^2)}}$, $\mbox{cos}\phi_R = \frac{h_{Rx}}{\sqrt{(d_{RD}^2+h_{Rx}^2)}}$ and $\mbox{sin}\gamma_R = \frac{d_{SR}}{\sqrt{(d_{SR}^2+h_{Tx}^2)}}$. Using above, the total received optical power, $P_{Rx}$ from all reflected paths at $D$ can be expressed as \cite{c2c}:
\begin{equation}\label{p_road}
    P_{Rx} =  \iint_{S} dP_{Rx}\,dS = \iint_S \frac{\mu' I(\zeta_R,\,\xi_R) \,d_{SR}^{-1} \, d_{RD}^{-1} \,h_{Rx} \;dS}{ (d^2_{SR}+h^2_{Tx})^\frac{1}{2}.(d^2_{RD}+h^2_{Rx})},
\end{equation}

where $\mu' = \frac{\rho_R A_R}{LER \,\pi}$. Utilizing \eqref{p_road}, BER can be calculated using \eqref{ber}.

% The non line of sight channel gain is expressed as \cite{vlc}:
% \begin{equation}\label{eq4}
%         H_{NLOS}=
%         \left\{ \begin{array}{ll}
%           \frac{K(m+1)}{2\pi D^2_{1} D^2_{2} } \cos^m(\phi)\cos(\alpha) \cos(\beta), & 0\leq\Psi\leq\Psi_{c}\\
%           0, & \text{otherwise}
%         \end{array} \right.
% \end{equation}

% where $K = \rho_R A_R T(\Psi) g(\Psi)$, $\rho_R$ is the reflection coefficient of the vehicle's side walls. $\Phi_{\frac{1}{2}}$ is the radiation angle at half-intensity. $A_R$ is the area of the photodetector at the receiver, $\Psi$ , $\Phi$, $T(\Psi)$ and $g(\Psi)$ are the the angle of irradiance, the angle of incidence, optical filter gain and the non-imaging concentrator, respectively. $g(\Phi)$ can be expressed as $g(\psi)= \frac{n^2}{\mathrm{sin^2(\Psi_{c})}}, \;;  0\leq\Psi\leq\Psi_{c},$, where $\Psi_c$ is the field of view (FOV) of the PD and $\n$ is the refractive index of the optical concentrator. $D$ is the distance between the source vehicle $S$ and the destination vehicle $D$. $D_1$ and $D_2$ denote the distances covered by the signal from $S$ to the road, and the road to $D$, respectively, as shown in Fig. \ref{fig_nlos}. Here $\alpha$ and $\beta$ are the incidence and reflectance angle w.r.t normal to reflecting surface.

\section{{Results and Discussions}}
In this section, we present the numerical results obtained through proposed scheme and compare the performance results with baseline scenarios such as optical relay and NRR aided V2V communication. The O-IRS consists of $n_m \times n_m$ mirror elements with dimension $0.01 \times 0.01 \,\mbox{m}^2$. The key simulation parameters adopted according to practical vehicular scenario are summarized in Table \ref{tab:chn}. All the presented results have been corroborated through Monte Carlo simulations. Unless otherwise stated, we assume distance of the destination vehicle, $d_{RD}$ to be $10\,\mbox{m}$.

Fig. \ref{fig_power_dist} shows received power as a function of distance, $d_{SR}$ for proposed O-IRS scheme and baseline scenarios. As intuitive, the received power decreases with increase in distance, $d_{SR}$. In particular, the received power for O-IRS aided V2V is more as compared to NRR aided V2V. Irrespective of any distance, $d_{SR}$, the received power for optical relay assisted V2V is more than the $5\times5$ O-IRS. However, $50\times50$  O-IRS system outperforms the optical relay aided V2V system. Thus, the number of mirror elements,  $n_m$ plays a key role in determining the performance of such O-IRS aided V2V system. Compared to an optical relay aided V2V,  an improvement of 15 dB and 30 dB in the received power can be observed for $20\times20$ and $50\times50$ O-IRS scheme respectively. Next, we also examine the impact of $n_m$ on received power. 

The variation of received power with increasing number of IRS elements, $n_m$ for two distinct values of distance of $S$ from O-IRS, $d_{SR}\in\{100\,\mbox{m}, 200\,\mbox{m}\}$ has been shown in Fig. \ref{fig_power_nm}. It is evident that the optical relay exhibit better performance as compared to O-IRS aided V2V with $n_m$$\leq$10. However, with increase in the number of mirror elements, the O-IRS system achieves substantially better received power than the optical relay. Compared to the O-IRS and optical relay aided V2V,  NRR aided V2V system offers very low received power.

 For the purpose of visualization and gain more insights, we also plot a 3D curve illustrating the cumulative impact of $d_{SR}$ and $n_m$ on received power for O-IRS and optical relay aided V2V system in Fig. \ref{fig_3d}. Interestingly, O-IRS performs better as compared to optical relay for the non red-colored regions (right to the black solid line) as depicted in Fig. \ref{fig_topview}. Irrespective of distance, $d_{SR}$, the proposed O-IRS scheme has more received power as compared to optical relay aided V2V for $n_m >$ 9.

 \begin{figure*}[t]
\centering
  \subfigure[BER versus $d_{SR}$ for O-IRS, Optical Relay and NRR, for different values of $n_m$.]{%
  \includegraphics[width=0.46\textwidth]{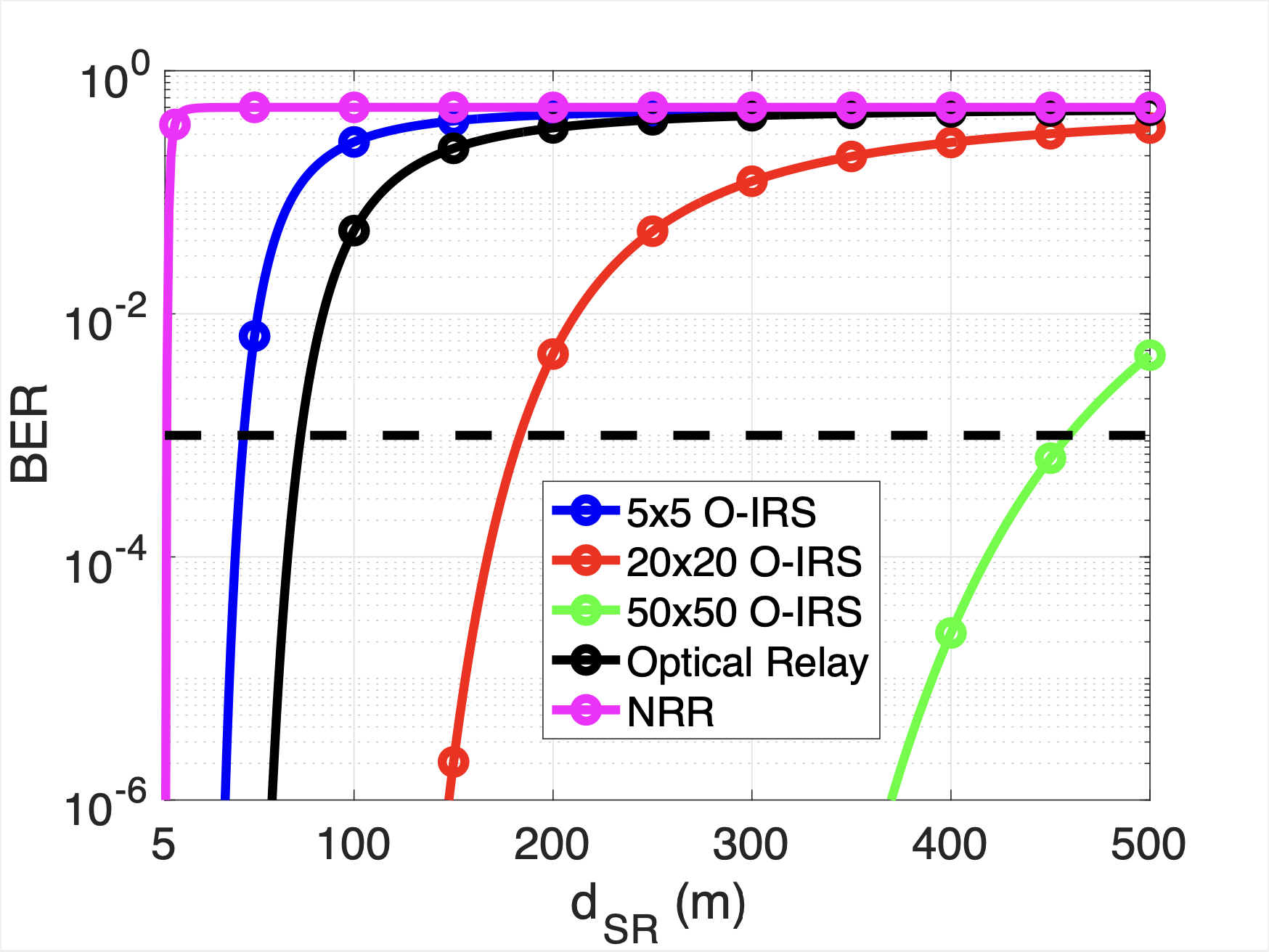}
 \label{ber_dist}}
 \subfigure[Impact of $n_m$ on BER, for different values of $d_{SR}$.]{%
 \includegraphics[width=0.45\textwidth]{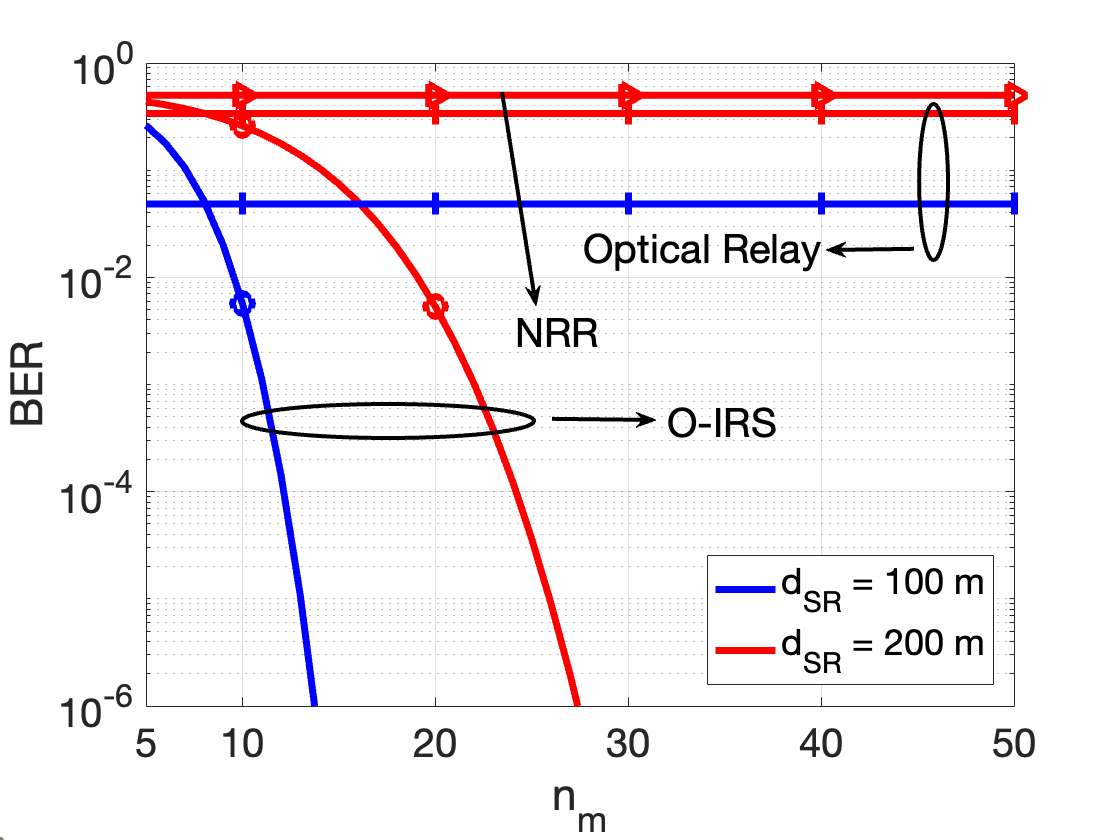}
 \label{ber_nm}}
 \caption{BER performance for the proposed scheme and the baseline scenarios.}
 \label{fig_ber}
 \end{figure*}

\begin{table}[t]
\centering
\caption{Maximum allowable, $d_{SR}$ to achieve a BER of $1\times10^{-3}$ for $n_m \times n_m$ O-IRS and PMS.\\}
\label{tab:ber_dist}
\begin{tabular}{|c|c|c}
\hline
\textbf{System} & \textbf{Required $d_{SR}$}  \\ \hline
NLOS Road Reflections & 6 m \\
\hline
Optical Relay & 73 m \\
\hline
$5\times5$ O-IRS & 45 m \\
\hline
$20\times20$ O-IRS & 184 m \\
\hline
$50\times50$ O-IRS & 459 m \\
\hline
\end{tabular}
\end{table}

We also plot BER performance of proposed scheme and compare with NRR and optical relay aided V2V communication. For critical safety applications, it needs to be ensured that the data transmission occurs with minimal error. The impact of increasing $d_{SR}$ on BER performance of proposed scheme is presented in Fig \ref{ber_dist}. Irrespective of any distance, $d_{SR}$, the proposed scheme offers better BER performance in comparison to NRR aided V2V. A further improvement in BER performance can be observed by increasing the number of mirror elements, $n_m$. Notice that unlike NRR aided V2V system, for a given BER performance, O-IRS aided V2V allows the source vehicle to transmit data from a larger communication distance. With BER of $1\times10^{-3}$ as performance benchmark, the performance of proposed scheme and other baseline scenarios has been summarized in Table \ref{tab:ber_dist}. It can be noted that for larger values, say for instance $n_m$=50, O-IRS significantly outperforms optical relay aided V2V in terms of allowable source vehicle’s distance while ensuring desired QoS requirement. 
For sake of completeness, we also show the impact of $n_m$ in the O-IRS on the BER performance for two different values of transmitter’s location, $d_{SR}\in\{100\,\mbox{m}, 200\,\mbox{m}\}$ as shown in Fig. \ref{ber_nm}. It can be observed that the NRR aided V2V does not offer reliable option to establish a communication link between the vehicles. It can be seen from the result that the increase in $n_m$ significantly improves the BER performance which is considerable better when compared to optical relay aided V2V. 
\section{Concluding Remarks}
In this work, we proposed a novel scheme employing optical IRS at the road intersections to facilitate V-VLC for road safety applications. From obtained results, it
has been shown that the proposed O-IRS outperforms the baseline scenarios such as optical relay and NRR aided vehicular communication systems. For a given number of mirror elements, $n_m$, O-IRS outperforms optical relay aided V2V in terms of maximum allowable source vehicle’s distance while ensuring desired QoS requirement. Additionally, the results also show that for a given source vehicle's distance, $d_{SR}$ and BER of $1\times10^{-3}$ as performance constraint, O-IRS with $n_m$ = 50, performs the significantly better as compared to optical relay and NRR aided V2V. Nonetheless, several distinctive
research challenges such as channel estimation in highly
dynamic environment, ambient sunlight impact, reliable energy management schemes, optimal
resource allocation and reflection optimization have to be
carefully addressed before the practical integration of O-IRS
into vehicular communication systems.

\bibliographystyle{IEEEtran}
\bibliography{asrj5}
\end{document}